\documentclass[10pt, prd, aps, amssymb, amsmath, tightenlines, noshowkeys, twocolumn]{revtex4}

\usepackage{graphicx}
\usepackage[T1]{fontenc}
\usepackage[latin1]{inputenc}
\usepackage[english]{babel}
\usepackage{amsmath}
\usepackage{amsfonts}
\usepackage{amssymb}
\usepackage{amsthm}
\usepackage{dsfont}
\usepackage{color}
\usepackage{xcolor}
\usepackage{float}
\usepackage{afterpage}
\usepackage{slashed}
\date{\today}

\begin{document}

\title{Continuous quantum phase transition in the fermionic mass solutions of the Nambu-Jona-Lasinio model}

\author{Alireza Beygi$^1$}
\email{beygi@thphys.uni-heidelberg.de}
\author{S. P. Klevansky$^1$}
\email{spk@physik.uni-heidelberg.de}
\author{R. H. Lemmer$^2$}
\email{deceased}
\affiliation{$^1$Institut f\"{u}r Theoretische Physik, Universit\"{a}t Heidelberg, Philosophenweg 12, 69120 Heidelberg, Germany\\
$^2$Physics Department, University of the Witwatersrand, Johannesburg, South Africa}

\begin{abstract}

Recently quantum simulators have been constructed to investigate experimentally the most prominent theoretical four-point many-body system described by the Hubbard model. By varying the coupling strength of the four-point interaction in relation to the kinetic term, it is possible to analyze the phase structure of the model. This intriguing fact leads us to ask the question as to whether similar Hamiltonians with four-point interactions can also be studied as a function of their four-point coupling strength. In this paper, we reexamine the Nambu-Jona-Lasinio model, regarding it generally beyond the context of quantum chromodynamics. Essentially, it is a model in which particle-antiparticle pairing leads to a BCS-like condensate, with the result that chiral symmetry is broken dynamically in the strong-coupling regime, where $G \Lambda^2$ is larger than a critical value, i.e., $G \Lambda^2 > G_c \Lambda^2$. To study the behavior of the system, it is necessary to move from this regime to a hypothetical regime of weak coupling, altering the coupling strength of the interaction arbitrarily. In order to do this, the gap equation must be regarded as complex and its Riemann surface structure must be known. We do this and obtain a continuous quantum phase transition characterized by the development of a complex order parameter (the dynamically generated mass) from the second sheet of the Riemann surface associated with the gap equation, as we move into the weak-coupling regime. The power-law behavior of the order parameter in the vicinity of the phase transition point is demonstrated to be independent of the choice of the regularization scheme with the critical exponent as $\beta \approx 0.55$. At the same time, the isovector pseudoscalar modes retain their feature as Goldstone modes and still have zero mass, while the isoscalar scalar meson follows the behavior of the order parameter and gains a width. Energetically, this mode is not favored over the normal, uncondensed mode but would have to be accessed through an excitation process.

\end{abstract}

\maketitle

\section{Introduction}
\label{s1}
With the turn of the century, there has been a revolution in quantum mechanics in that new tools have become available to explore and control quantum systems and their dynamics, either through the construction and manipulation of synthetic systems or of natural ones through the use of new materials. Especially in cold atomic systems, methods have been developed to simulate otherwise difficult-to-solve many-body problems \cite{bbb11, www, 20}. One prominent example is the construction of a quantum simulator for the Hubbard model. This model contains four-point interactions within the Hamiltonian
\begin{equation}
{H}_{\textrm{Hubbard}} = -t \sum_{i,j,\alpha} \hat c^\dagger_{i\alpha} \hat c_{j\alpha} + \frac U2 \sum_{i, \alpha\ne \beta} \hat n_{i\alpha} \hat n_{i\beta},
\label{eq:1}
\end{equation}
that is expressed in second-quantized form. The operator $\hat c_{i\alpha}$ ($\hat c^\dagger_{i\alpha}$) destroys (creates) a particle at site $i$ with quantum number $\alpha$, so that $\hat n_{i\alpha} = \hat c^\dagger_{i\alpha} \hat c_{i\alpha}$ counts the number of particles at site $i$ and $U$ is the interaction strength. Experimentally it is now possible to adjust the coupling strength $U$ in relation to the kinetic energy in order to observe and quantify a possible phase transition.  Adjusting the coupling, for example, has allowed for an experimental observation of the  BEC-BCS crossover \cite{3,4,5,6,29}. These experiments are based on the fact that particles can be trapped in optical lattices and be manipulated to high precision.

It is tantalizing to hope that cold atomic physics may one day provide deeper insights into the phase transitions of quantum chromodynamics (QCD). The chiral phase transition, in particular, can also be well-modeled by a Hamiltonian that contains a four-point interaction, similar to that in (\ref{eq:1}). This is the Nambu-Jona-Lasinio (NJL) Hamiltonian density \cite{SPK1},
\begin{equation}
{\cal H}_{\textrm{NJL}} = \bar \psi (\boldsymbol{\gamma} \cdot {\bf p} + m_0) \psi 
 - G[(\bar\psi\psi)^2 + (\bar\psi i\gamma_5 \tau\psi)^2].
\label{eq:2}
\end{equation}
In this expression, the first term expresses the (relativistic) kinetic energy, with $\boldsymbol{\gamma}$ denoting the 4-dimensional Dirac matrices; $m_0$ denotes a current mass, and the interaction strength is given as $G$, the minus sign being purely a convention. Two interaction terms  $n_S^2$ and $n_{PS}^2$, with $n_S =  \bar\psi\psi$ and $n_{PS}= \bar\psi i\gamma_5 \tau\psi$, where $\tau$'s represent the isospin SU(2) matrices, are necessary in order to preserve the chiral symmetry of the interaction for the two-flavor version of the model. 

Within this effective field theory one can study how the mechanism of chiral symmetry breaking functions within a theory of interacting fermions \cite{SPK1}. It does so in a way that parallels the mechanism of pairing in the Bardeen-Cooper-Schrieffer (BCS) theory of superconductivity \cite{100}. In the BCS theory, pairing takes place between like particles, that is, electrons with opposite spins. In the NJL model, the pairing takes place between particles and their antiparticles, that is, between fermions and antifermions. This can be quantified by constructing a trial ground state 
\begin{equation}
| 0\rangle_{\textrm{NJL}} = \prod_{{\bold p}, s=\pm1} [\cos\theta (p) + s \sin \theta(p) b^\dagger({\bold p},s) d^\dagger (-{\bold p},s)] |0\rangle,
\label{eq:3}
\end{equation}
in which a variational parameter $\theta(p)$ is introduced to measure the strength of the pairing of a fermion with momentum ${\bold p}$ and helicity $s$ with an antifermion of opposite momentum $-{\bold p}$ but also helicity $s$, relative to the ground state of the associated basis $|0\rangle$, defined through $b({\bold p},s)|0\rangle = d({\bold p},s)|0\rangle = 0$. Minimizing the ground state energy ${}_{{\textrm{NJL}}}\langle 0|{\cal H}_{\textrm{NJL}} |0\rangle_{\textrm{NJL}}$ leads to the gap equation
\begin{equation}
p \tan 2\theta (p) = 4 G N_cN_f \int \frac{d^3q}{(4\pi)^3} \sin 2\theta (q),
\label{eq:3.01}
\end{equation}
where $N_c$ is the number of colors and $N_f$ the number of flavors of the system. From (\ref{eq:3.01}), one can deduce that $\theta (p)$ is independent of $p$ and that for certain values of the coupling $G$ the value of $\theta$  is non-zero. The identification of $\tan 2\theta (p) = m^*/p$ completes the argument and leads to the well-known form
\begin{equation}
m^* = 4GN_cN_f \int \,\frac{d^3p}{(2\pi)^3}\frac {m^*}{E_p}
\label{eq:3.1}
\end{equation}
for the gap equation, which is regulated with an $O(3)$ cut-off.

A more direct comparison of (\ref{eq:2}) with (\ref{eq:1}) can be made by expanding the field operators in terms of the second-quantized creation and annihilation operators,
\begin{eqnarray}
\psi({\bf x})& =& \sum_s \int d{\tilde p} [ b ({\bold p},s)u ({\bold p},s) + d^\dagger(-{\bold p},s)v (-{\bold p},s)]e^{i{\bold p}\cdot {\bold x}}, \nonumber \\
\bar \psi ({\bf x})& =& \sum_s \int d{\tilde p}  [b^\dagger ({\bold p},s)\bar u ({\bold p},s) 
 + d (-{\bold p},s)\bar v (-{\bold p},s)]e^{-i{\bold p}\cdot {\bold x}}, \nonumber \\
\label{eq:4}
\end{eqnarray}
with $d{\tilde p} = [d^3p/(2\pi)^3](m_0/E_p)$. Then the kinetic term of (\ref{eq:2}), integrated over ${\bold x}$, takes the form
\begin{equation}
 T= \sum_s\int \, d\tilde p E_p[ b^\dagger ({\bold p},s) b ({\bold p},s) + d^\dagger ({\bold p},s) d ({\bold p},s)]. 
\label{eq:5}
\end{equation}
The comparison with the kinetic term in (\ref{eq:1}) is evident: The NJL model has two species of particles. The interaction terms that are possible are correspondingly more involved.

The general starting point for understanding the phase structure of the interacting fermionic system is the gap equation. A more generalizable form of it follows from field-theoretic considerations, by identifying the self-consistent self-energy $\Sigma (x,y)$ in the mean-field approximation that arises from the four-point interactions in (\ref{eq:2}),
\begin{equation}
\Sigma(x,x) =  2iG{\rm Tr} S(x,x),
\label{eq:6}
\end{equation}
where Tr is the trace over all degrees of freedom and $S(x,y)$ is the self-consistent Green function defined through
\begin{equation}
 [i\slashed \partial_x - \Sigma (x,y)] S(x,y) = \delta^{(4)} (x-y).
\label{eq:6.1}
\end{equation}
Extensions of the Lagrangian and the resulting gap equation to include the effects of external parameters such as temperature or external electromagnetic fields on the phase diagram can be studied \cite{SPK2} and there is an extensive literature on the NJL model, expecially in this context.

The variation of such external parameters usually follows once the model parameters of the NJL Lagrangian, i.e., the interaction coupling strength $G$ and a regulatory cut-off $\Lambda$, have been fixed. The fact that different regularization schemes that introduce $\Lambda$ all lead to values of the coupling where $G \Lambda^2 > G_c \Lambda^2$, where $G_c$ is some critical value of the coupling strength, reinforces the model as a \textit{strong}-coupling theory, akin to QCD, and leads to fermionic quasiparticle masses which can be identified as dressed or constituent quark masses.

However, as with most many-body theories, these calculations are approximate: The gap equation (derived simply through the energy argument above or formally through diagrammatic methods giving rise to the self-energy in terms of the Green function as in (\ref{eq:6})) corresponds to the self-consistent mean-field approximation. This fact in itself would render measurements from a quantum simulator containing two different species with corresponding interactions extremely useful.

The question that is addressed in this paper, however, goes back to a basic, if for the moment, only theoretical question.  In analogy to the questions posed in understanding the Hubbard model and analogies in describing the BEC-BCS crossover \cite{3, 4, 5, 6, 29}, we seek to understand what happens when the interaction coupling of a system of fermions interacting via an NJL-type Lagrangian is altered to such an extent that one moves into the \textit{weak}-coupling regime of the theory: Instead of fixing the NJL coupling strength to its usual regularization dependent strong-coupling value, we treat it as a parameter and look for the solution of the gap equation as a function of this parameter. The difficulty in this lies in the fact that the relevant equations, which up to now have always been treated as having real variables, must be regarded as complex.  In what follows, we keep the QCD-notation of the NJL model in order to check the validity of our results on the real axis, but we abstract from this in thought in regarding the model as a two-component fermionic model with specific interaction. 

We find a continuous quantum phase transition characterized by the development of a width for the dynamically generated fermion mass onto the higher sheets of the Riemann surface associated with the gap equation. By an appropriate choice of the order parameter, we show that the power-law behavior of the phase transition does not depend on our choice of regularization scheme - we have demonstrated this with the covariant, Pauli-Villars, and proper-time schemes. The value of the mass of the Goldstone particle is unaffected by this transition, however, the mass of the associated scalar meson also develops an imaginary part.

This paper is organized as follows: In Sec.~\ref{ss11} we discuss the Riemann surface structure associated with the gap equation. In Sec.~\ref{ss2} we solve the gap equation for its spectrum in the covariant regularization scheme, in both the strong- and weak-coupling regimes. We examine the stability of solutions against the choice of the regularization scheme by studying the gap equation in the context of the Pauli-Villars regularization scheme in Sec.~\ref{ss3}, and we observe the same behavior of the order parameter in the vicinity of the phase transition point as it is obtained in the covariant regularization scheme. This is again verified in Sec.~\ref{ss4}, using the proper-time regularization scheme, and extended to include the effects of incorporating a constant electric field. Then in Sec.~\ref{s4} we comment on the effects of the phase transition on the associated isovector pseudoscalar and isoscalar scalar modes. We summarize and conclude in Sec.~\ref{s6}.

%With the turn of the century, there has been a revolution in quantum mechanics in that the construction of synthetic %quantum systems has become possible. Especially in cold atomic and molecular physics, in some prepared cases, the %coupling strength of the interaction itself can be adjusted \cite{bbb11, www, 20}.

\section{Weak-coupling fermionic mass solutions}
\label{ss11}

\subsection{Solutions of the gap equation in the covariant regularization scheme}
\label{ss2}

\begin{figure}
\centering
\includegraphics[scale = 0.45]{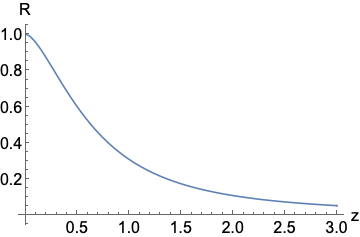}
\caption{At $z = 0$, the right-hand side of (\ref{cogap}) reaches its maximum of $1$.}
\label{conew}
\end{figure}

Since the interaction terms in (\ref{eq:2}) are point-like, the self-energy in (\ref{eq:6}) is constant and is thus identified as the dynamically generated mass, $\Sigma = m^*$. Thus, the solution to the Green function equation containing $\Sigma$, Eq.~(\ref{eq:6.1}), is simple: In momentum space it is $S(p) = (\slashed p + m^*)/(p^2-m^{*2})$, which can be inserted into (\ref{eq:6}). The integral arising on the right-hand side of the gap equation,
\begin{equation}
\int \frac{d^4p}{(2\pi)^4} {\rm Tr}  S(p),
\label{eq:7}
\end{equation}
diverges and must be regulated. $O(3)$ regularization leads to (\ref{eq:3.1}). In the covariant regularization scheme, which we will consider further here, the Euclidean four-momentum is restricted, $p_E^2 = {\bf p}^2 +p_4^2\le \Lambda^2$, where $p_0=ip_4$. Consequently the gap equation takes the form
\begin{equation}
m^* = \frac{1}{2 \pi^2} N_c N_f G \Lambda^2 m^* \Big[1 - \frac{m^{*2}}{\Lambda^2} \ln\Big (1 + \frac{\Lambda^2}{m^{*2}} \Big)\Big].
\end{equation}
Canceling the $m^*$ on both sides, one has the well-known result \cite{SPK1},
\begin{equation}
\frac{2 \pi^2} {N_c N_f G \Lambda^2} = 1 - z^2 \ln\Big[1 + \frac 1{ z^2}\Big],
\label{cogap}
\end{equation}
where $z = m^* / \Lambda$. In order to obtain a real solution for $m^*$, the left-hand side of (\ref{cogap}) should be less than one. (The right-hand side of (\ref{cogap}), denoted as $R(z)$, has a global maximum of $1$ at $z =0$, see 
Fig.~\ref{conew}.) This leads to $2 \pi^2 / (N_c N_f) = \pi^2 / 3 < G \Lambda^2$, where $N_cN_f=6$. Thus, the usual real solution for $m^*$ lies in the strong-coupling regime, where $2 \pi^2 / (N_c N_f) \equiv G_c \Lambda^2 \approx 3.29$ is the critical value of the coupling strength.

The standard choices for the regulatory cut-off as $\Lambda = 1015$ MeV and the coupling strength as $G \Lambda^2 = 3.93$ \cite{SPK1} satisfy this inequality. With these parameters, we obtain for $m^*$ the value $238.486$ MeV, which gives a good estimate for a constituent quark mass.

\begin{figure}
\centering
\includegraphics[scale = 0.4]{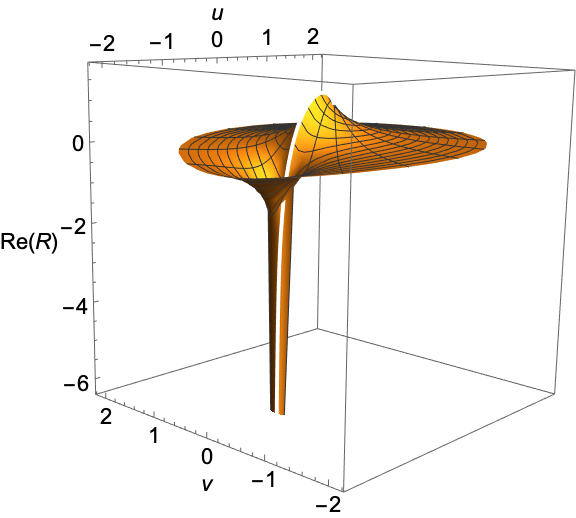}
\caption{The real part of the complex function $R(w)$. The branch cut is shown as a white line from $-1 \to 0$ along the $u$-axis.}
\label{cut}
\end{figure}

The objective here is to generalize the real gap equation to the complex plane through treating the coupling strength $G$ as a variable and to look for the solutions of the gap equation as a function of $G$. To investigate the solutions of (\ref{cogap}) for arbitrary values of $G$, it is convenient to denote the right-hand side of (\ref{cogap}) as $R(w)$, where $z^2 \equiv w = u + i v$, yielding
\begin{equation}
R(w) = 1 - w \ln[(1 + w) / w].
\label{R10}
\end{equation}
We set
\begin{equation}
\begin{split}
&w = |w| e^{i \phi} \quad \quad 0 < \phi < 2 \pi, \\
&1 + w = |1 + w| e^{i \theta} \quad \quad 0 < \theta < 2 \pi,
\end{split}
\label{R20}
\end{equation}
where $|w| = \sqrt{u^2 + v^2}$, $|1 + w| = \sqrt{(1 + u)^2 + v^2}$, $\tan\phi = v / u$, and $\tan\theta = v / (1 + u)$. Equations (\ref{R10}) and (\ref{R20}) define a branch cut from $-1 \to 0$ along the $u$-axis. Now, by varying the two angles $\phi$ and $\theta$, we can traverse the complex plane. The cut-plane for the function $R(w)$ is shown in Fig.~\ref{cut}.
%\begin{figure}
%\centering
%\includegraphics[scale = 0.36]{100}
%\caption{The left panel shows the plot of real part of the complex function $R(w)$, which shows the branch cut, as a %white line, from $-1 \to 0$ along the $u$-axis. The right panel demonstrates the contour lines of the same function.}
%\label{cut}
%\end{figure}

By exploiting (\ref{R20}), (\ref{R10}) can be written as
\begin{equation}
R = 1 - |w| e^{i \phi} \Big[\ln\big[|1 + w| / |w|\big] + i (\theta - \phi)\Big].
\label{NJLCO}
\end{equation}

\begin{figure}[h]
\centering
\includegraphics[scale = 0.18]{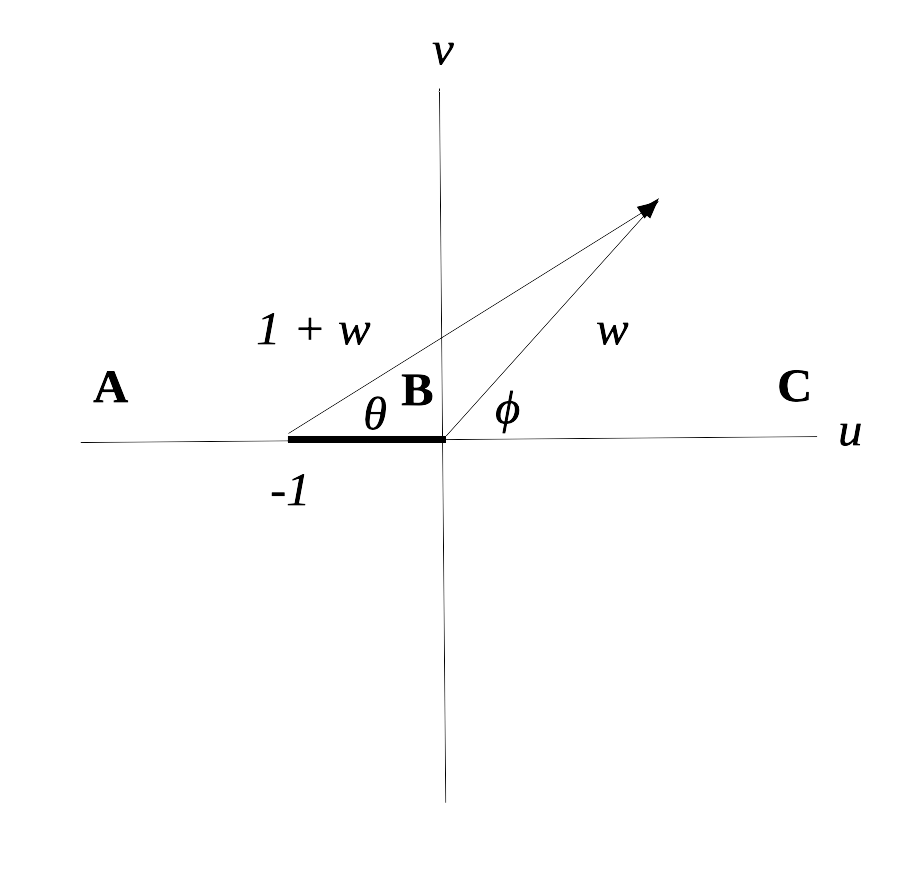}
\caption{In the first sheet of the Riemann surface of $R(w)$, we have: $0 < (\phi, \theta) < 2 \pi$. The branch cut from $-1 \to 0$ is shown by a thick line.}
\label{w1}
\end{figure}

The first sheet of the Riemann surface of $R(w)$ is defined by restricting the angles to $0 < (\phi, \theta) < 2 \pi$. To demonstrate the discontinuity along the branch cut, first we identify three regions, namely, \textit{A}: $-\infty < u < -1$, \textit{B}: $-1 < u <0$, and \textit{C}: $0 < u < +\infty$, as depicted in Fig.~\ref{w1}.
\\
Along the upper lip, i.e., $v \to 0^+$, in the region \textit{A}, we have:
\begin{equation}
-\infty < u < -1 \quad \quad \phi = \theta = \pi,
\end{equation}
for which, according to (\ref{NJLCO}),
\begin{equation}
R = 1 + |u| \ln\big[|1 + u| / |u|\big].
\label{R}
\end{equation}
Along the lower lip, i.e., $v \to 0^-$, the angles $\phi$ and $\theta$ are the same:
\begin{equation*}
-\infty < u < -1 \quad \quad \phi = \theta = \pi,
\end{equation*}
as a result, $R$ has the same form as (\ref{R}). We conclude that the complex function $R(w)$ is continuous for $-\infty < u < -1$.
\\
For the stretch \textit{B}, along the upper lip, the angles are
\begin{equation}
-1 < u < 0 \quad \quad \phi = \pi \quad \quad \theta = 0,
\end{equation}
which results in
\begin{equation}
R = 1 + |u| \Big[\ln\big[|1 + u| / |u|\big] - i \pi\Big].
\label{R1}
\end{equation}
However, along the lower lip, we have:
\begin{equation}
-1 < u < 0 \quad \quad \phi = \pi \quad \quad \theta = 2 \pi,
\end{equation}
which leads to
\begin{equation}
R = 1 + |u| \Big[\ln\big[|1 + u| / |u|\big] + i \pi\Big].
\label{R2}
\end{equation}
Equations (\ref{R1}) and (\ref{R2}) show the discontinuity between the upper- and lower-lip values of $R(w)$ for $-1 < u <0$ explicitly, as is expected by crossing the branch cut. In region $C$, the function $R$ is again continuous. This can easily be checked by noting that along the upper lip of the region \textit{C}, the angles are
\begin{equation}
0 < u < +\infty \quad \quad \phi = \theta = 0,
\end{equation}
for which
\begin{equation}
R = 1 - |u| \ln\big[|1 + u| / |u|\big],
\label{RC}
\end{equation}
while along the lower lip the angles become
\begin{equation}
0 < u < +\infty \quad \quad \phi = \theta = 2 \pi,
\end{equation}
which results in the same $R$ as (\ref{RC}).

We now examine the second sheet of the Riemann surface of the complex function $R(w)$, which is defined by $0 < \phi < 2 \pi$ and $-2 \pi < \theta < 0$. The same analysis as above reveals for region A (continuous):
\begin{equation}
\begin{split}
&-\infty < u < -1 \quad v \to 0^+: \\
&\phi = \pi \quad \theta = -\pi \quad R = 1 + |u| \Big[\ln\big[|1 + u| / |u|\big] - 2 \pi i \Big],
\end{split}
\end{equation}
\begin{equation}
\begin{split}
&-\infty < u < -1 \quad v \to 0^-: \\
&\phi = \pi \quad \theta = -\pi \quad R = 1 + |u| \Big[\ln\big[|1 + u| / |u|\big] - 2 \pi i \Big],
\end{split}
\end{equation}
region B (discontinuous):
\begin{equation}
\begin{split}
&-1 < u < 0 \quad v \to 0^+: \\
&\phi = \pi \quad \theta = -2 \pi \quad R = 1 + |u| \Big[\ln\big[|1 + u| / |u|\big] - 3 \pi i \Big],
\end{split}
\end{equation}
\begin{equation}
\begin{split}
&-1 < u < 0 \quad v \to 0^-: \\
&\phi = \pi \quad \theta = 0 \quad R = 1 + |u| \Big[\ln\big[|1 + u| / |u|\big] - i \pi \Big],
\end{split}
\label{CC}
\end{equation}
and region C (continuous):
\begin{equation}
\begin{split}
&0 < u < +\infty \quad v \to 0^+: \\
&\phi = 0 \quad \theta = -2 \pi \quad R = 1 - |u| \Big[\ln\big[|1 + u| / |u|\big] - 2 \pi i \Big],
\end{split}
\end{equation}
\begin{equation}
\begin{split}
&0 < u < +\infty \quad v \to 0^-: \\
&\phi = 2 \pi \quad \theta = 0 \quad R = 1 - |u| \Big[\ln\big[|1 + u| / |u|\big] - 2 \pi i \Big].
\end{split}
\end{equation}
As on the first sheet, there is a discontinuity between the upper- and lower-lip values of $R(w)$ for $-1 < u <0$.

\begin{figure}
\centering
\includegraphics[scale = 0.32]{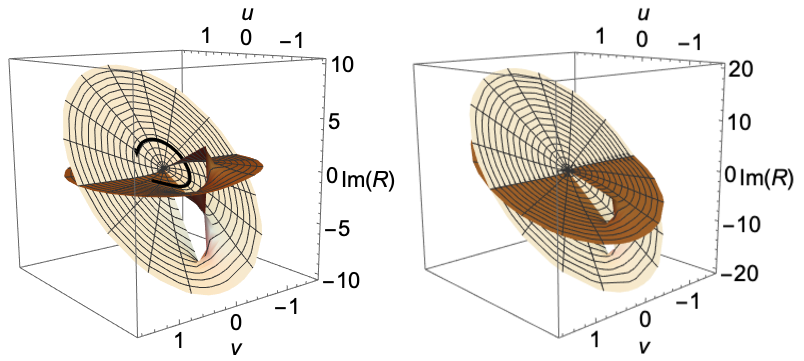}
\caption{The Riemann surface of the complex function $R(w)$. In the left panel, the first two sheets are connected along the branch cut from $-1 \to 0$ along the $u$-axis. By encircling one of the branch points and continuously crossing the branch cut from positive $v$ to negative $v$ (see the thick black line), we switch between the sheets. The second and the third sheets are shown in the right panel. On the third sheet we have: $0 < \phi < 2 \pi$ and $-4 \pi < \theta < -2 \pi$.}
\label{RR}
\end{figure}

Equations (\ref{R1}) and (\ref{CC}) demonstrate the continuous join of the first sheet upper-lip value with the second sheet lower-lip value of $R(w)$ along the branch cut from $-1 \to 0$ along the $u$-axis; in other words, by encircling the branch point and continuously crossing the branch cut from $v > 0$ to $v < 0$, we move to the second sheet, see Fig.~\ref{RR}.
%\begin{figure}
%\centering
%\includegraphics[scale = 0.36]{R}
%\caption{The Riemann surface of the complex function $R(w)$. The first two sheets are connected along the branch cut %from $-1 \to 0$ along the $u$-axis. By encircling one of the branch points and continuously crossing the branch cut %from positive $v$ to negative $v$ (see the right panel), we switch between the sheets.}
%\label{RR}
%\end{figure}

\subsubsection{Continuous quantum phase transition}

In the strong-coupling interaction domain where $G \Lambda^2 > G_c \Lambda^2 = 2 \pi^2 / (N_c N_f) = \pi^2 / 3$, by choosing, for example,  $G \Lambda^2 = 3.93$ \cite{SPK1}, the gap equation on the upper (or lower) lip of the strip $0 < u < +\infty$, Eq.~(\ref{RC}), reads
\begin{equation}
C = 0.8371 = 1 - |u| \ln\big[|1 + u| / |u|\big],
\end{equation}
where $C \equiv 2 \pi^2 / (N_c N_f G \Lambda^2) = G_c \Lambda^2 / (G \Lambda^2)$.
Here we find a single real root $u_r = 0.0552$, for which the constituent quark mass becomes $m^* = \Lambda \sqrt{u_r} = 238.471$ MeV, confirming the expected result. ($\Lambda$ has been set to $1015$ MeV.)

\begin{figure}
\centering
\includegraphics[scale = 0.6]{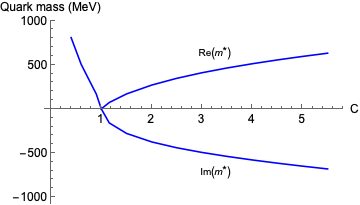}
\caption{The quark mass decreases as $C = G_c \Lambda^2 / (G \Lambda^2)$ increases. At the phase transition point, $C = 1$, by crossing the branch cut, the mass develops a width as one moves onto the second sheet of the Riemann surface. The mass scale has been set using $\Lambda = 1015$ MeV.}
\label{figure1}
\end{figure}

Now, if we treat $C$ (inverse of the coupling strength) as a parameter, by increasing $C$ (decreasing the coupling strength), the dynamically generated mass of the fermion decreases, see Fig.~\ref{figure1}, until it reaches a phase transition point, i.e., $C_c = G_c \Lambda^2 / (G \Lambda^2) = 1$, where by encircling the branch point, a width for the fermion mass is generated on the second sheet of the Riemann surface, that is, the mass becomes complex, see Fig.~\ref{figure1} and Table~\ref{table1}. We note that the phase transition in the vicinity of $C = 1$ is a continuous function of the system parameter, in this case the coupling strength.

By defining the order parameter to be the imaginary part of the mass, |Im($m^*$)|, we observe that it is zero before crossing the branch cut. However, by increasing $C$  through the phase transition point, the order parameter becomes non-zero on the second sheet of the Riemann surface, and it increases with increasing values of $C$.

\begin{figure}
\centering
\includegraphics[scale = 0.6]{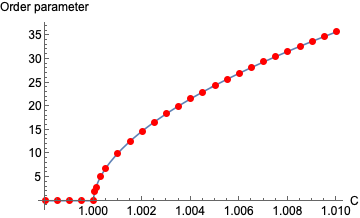}
\caption{Continuous quantum phase transition in the fermionic mass solution of the Nambu-Jona-Lasinio model. By choosing the order parameter as the imaginary part of the mass, the numerical data (red dots) indicate a power-law behavior in the vicinity of the phase transition point. By fitting a curve to the data we obtain the critical exponent to be approximately $0.55$: Order parameter $\propto (C - 1)^{0.55}$.}
\label{O}
\end{figure}

In the vicinity of the phase transition point, as illustrated in Fig.~\ref{O}, we fit a curve to the numerical data and find that the order parameter diverges with a power law:
\begin{equation}
 |\text{Im}(m^*)| \propto (C - 1)^\beta,
\end{equation}
where the critical exponent is found to be $\beta \approx 0.55$.

\begin{center}
\begin{table}
\begin{tabular}{|c|c|c|c|c|}
\hline
$G \Lambda^2$ & $C$ & $u_r$ & $v_r$ & $m^* = m - i \gamma$\\
\colrule
$\pi^2 / 3$ & $1.00$ & $0.0000$ & $0.0000$ & $0.0000 - 0.0000 i$\\

$2.84$ & $1.16$ & $-0.0195$ & $-0.0224$ & $72.4807 - 159.1940 i$\\

$2.19$ & $1.50$ & $-0.0471$ & $-0.0916$ & $169.6900 - 278.0620 i$\\

$1.64$ & $2.01$ & $-0.0654$ & $-0.1970$ & $270.6190 - 374.9820 i$\\

$1.32$ & $2.49$ & $-0.0719$ & $-0.2946$ & $345.2820 - 439.6510 i$\\

$1.10$ & $2.99$ & $-0.0736$ & $-0.3904$ & $408.3260 - 492.4980 i$\\

$0.94$ & $3.50$ & $-0.0728$ & $-0.4846$ & $463.5990 - 538.4470 i$\\

$0.82$ & $4.01$ & $-0.0709$ & $-0.5766$ & $512.5710 - 579.4590 i$\\

$0.73$ & $4.51$ & $-0.0684$ & $-0.6636$ & $555.3430 - 615.5270 i$\\

$0.66$ & $4.98$ & $-0.0658$ & $-0.7462$ & $593.2740 - 647.8910 i$\\
\hline
\end{tabular}
\caption{Quark masses, $m^*$'s, in the weak-coupling region, i.e., $C \geq 1$, for different $C$'s, on the second sheet of the Riemann surface which is defined by $0 < \phi < 2 \pi$ and $-2 \pi < \theta < 0$.}
\label{table1}
\end{table}
\end{center}

As a side remark, we note that the dynamically generated mass in the weak-coupling regime on the second sheet of the Riemann surface must be considered to be complex, i.e., $m^* = m - i \gamma$, so that the associated Feynman propagator, 
\begin{equation}
S(p) = \frac{1}{\slashed{p} - m + i \gamma},
\label{fey1}
\end{equation}
is used.

\subsection{Solutions of the gap equation in the Pauli-Villars regularization scheme}
\label{ss3}

\begin{figure}
\centering
\includegraphics[scale = 0.45]{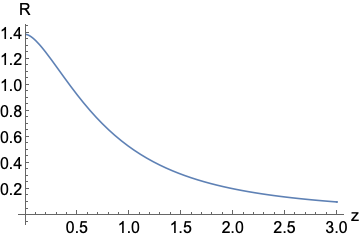}
\caption{The right-hand side of (\ref{RPV}) denoted as $R$ has a global maximum of $2 \ln2$.}
\label{RPV22}
\end{figure}

In order to investigate the stability of solutions of the gap equation against the choice of the regularization scheme, we study the gap equation again, now using the Pauli-Villars regularization scheme. Then Eq.~(\ref{cogap}) reads
\begin{equation}
2 \pi^2 / (N_c N_f G \Lambda^2) = F_2(z^2) / \Lambda^2,
\end{equation}
where $z = m^* / \Lambda$ and
\begin{equation}
\frac {F_2(z^2)}{  \Lambda^2} = \sum_{a = 0}^{2} C_a (\alpha_a + z^2) \ln\Big[1 + \frac{\alpha_a} { z^2}\Big],
\end{equation}
with $C_0 = 1$,  $\alpha_0 = 0$, $C_1 = 1$, $\alpha_1 = 2$, $C_2 = -2$, and $\alpha_2 = 1$. Then it follows that
\begin{equation}
\frac{2 \pi^2} {N_c N_f G \Lambda^2} = (2 + z^2) \ln[1 + 2 / z^2] - 2 (1 + z^2) \ln[1 + 1 / z^2].
\label{RPV}
\end{equation}
The right-hand side of (\ref{RPV}) has a global maximum of $2 \ln2$, as illustrated in Fig.~\ref{RPV22}. In order to obtain a real solution for the mass, the left-hand side of (\ref{RPV}) denoted as $C$ should be less than this maximum value; in other words,  $C \equiv 2 \pi^2 / (N_c N_f G \Lambda^2) < 2 \ln2$ or $\pi^2 / (6 \ln2) \equiv G_c \Lambda^2 \approx 2.37 < G \Lambda^2$, where $G_c \Lambda^2$ is the critical value of the coupling strength. As a result, the real solution for the mass again lies in the strong-coupling domain.  The usual choice of parameters for the regulatory cut-off and the coupling strength, $\Lambda = 859$ MeV and $G \Lambda^2 = 2.84$ \cite{SPK1}, satisfy the above requirement for a real mass, and leads to $m^* = 240.334$ MeV.

\begin{figure}
\centering
\includegraphics[scale = 0.42]{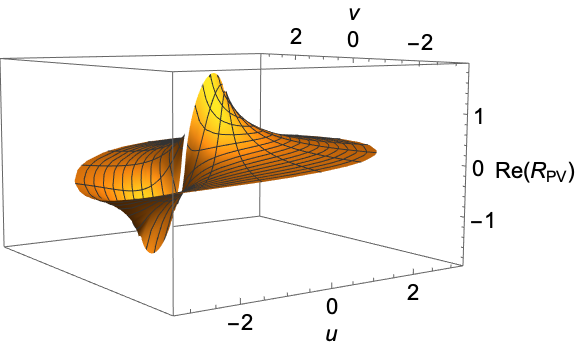}
\caption{The branch cut of the complex function $R_{\text{PV}}(w)$ is taken from $-2 \to 0$ along the $u$-axis.}
\label{BBBB}
\end{figure}

As in Sec.~\ref{ss2}, by moving to the complex plane and defining $z^2 \equiv w = u + i v$, we obtain the analog of the complex function $R(w)$ of Eq.~(\ref{R10}), which we denote as $R_{\text{PV}}(w)$:
\begin{equation}
R_{\text{PV}}(w) = (2 + w) \ln[2 + w] + w \ln w - 2 (1 + w) \ln[1 + w].
\label{RPVRPV}
\end{equation}
The first sheet of the Riemann surface of this complex function is defined via the three angles $\xi$, $\phi$, and $\theta$, where
\begin{equation}
\begin{split}
&2 + w = |2 + w| e^{i \xi} \quad \quad 0 < \xi < 2 \pi, \\
&w = |w| e^{i \phi} \quad \quad 0 < \phi < 2 \pi, \\
&1 + w = |1 + w| e^{i \theta} \quad \quad 0 < \theta < 2 \pi,
\end{split}
\label{PVVP}
\end{equation}
and $\phi$ and $\theta$ are defined as before: $\tan\phi = v / u$ and $\tan\theta = v / (1 + u)$. In the same way, for $\xi$, we have: $\tan\xi = v / (2 + u)$. We define the branch cut of $R_{\text{PV}}(w)$ from $-2 \to 0$ along the $u$-axis; this is  shown in Fig.~\ref{BBBB} as a white line.
We can now rewrite (\ref{RPVRPV}) using (\ref{PVVP}) as
\begin{equation}
\begin{split}
R_{\text{PV}} = &|2 + w| e^{i \xi} \big[\ln|2 + w| + i \xi\big]\\
&+ |w| e^{i \phi} \big[\ln|w| + i \phi\big]\\
&- 2 |1 + w| e^{i \theta} \big[\ln|1 + w| + i \theta\big].
\end{split}
\label{NJLPV1}
\end{equation}
By separating the real and imaginary parts of (\ref{NJLPV1}), we have:
\begin{equation}
\begin{split}
\text{Re}[R_{\text{PV}}(u, v)] = &(2 + u) \ln|2 + w| + u \ln|w|\\
&- 2 (1 + u) \ln|1 + w| - v (\xi + \phi - 2 \theta),
\end{split}
\label{GI}
\end{equation}
\begin{equation}
\begin{split}
\text{Im}[R_{\text{PV}}(u, v)] = &v \Big[\ln|2 + w| + \ln|w|\\
&-2 \ln|1 + w|\Big] + u (\xi + \phi - 2 \theta) + 2 (\xi - \theta).
\end{split}
\end{equation}

\begin{figure}
\centering
\includegraphics[scale = 0.18]{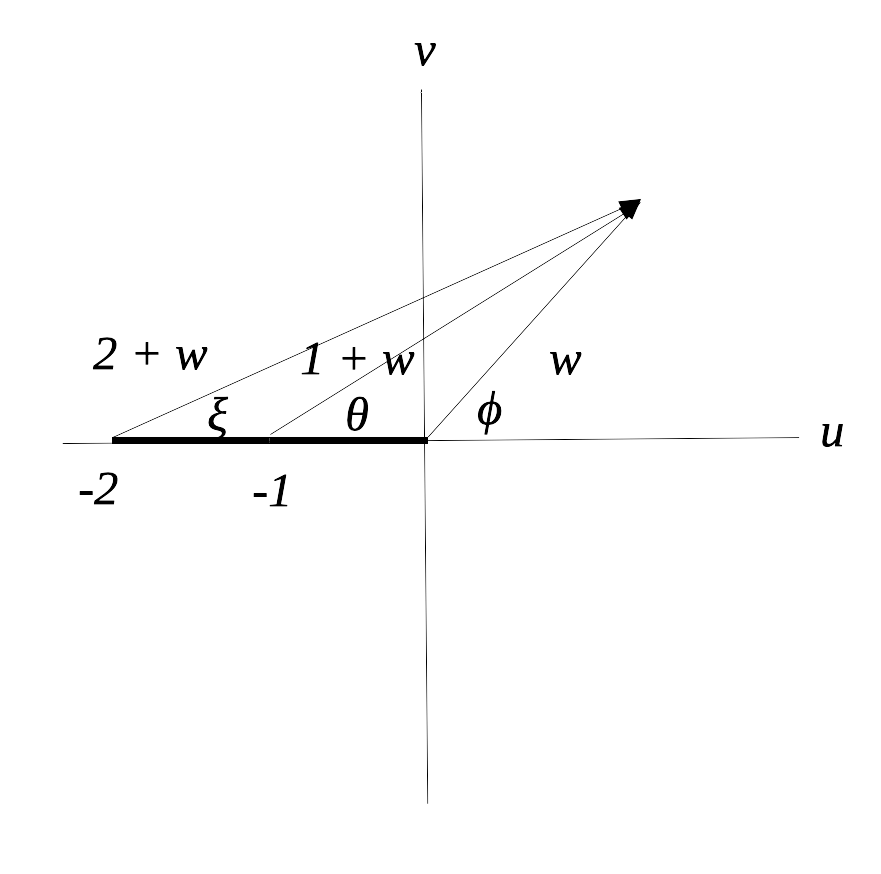}
\caption{Cut-plane for the complex function $R_{\text{PV}}(w)$. The first sheet of the Riemann surface of $R_{\text{PV}}(w)$ is defined by $0 < (\xi, \phi, \theta) < 2 \pi$.}
\label{newrpv}
\end{figure}

Next we examine the continuity of $R_{\text{PV}}$ along the upper/lower lip of the $u$-axis as $v \to 0^{\pm}$. The relevant angles for the sections connecting the branch points at $-2$, $-1$, and $0$, along the $u$-axis are shown in Fig.~\ref{newrpv}.
We analyze the behavior of $R_{\text{PV}}$ along four sections, namely, $-\infty < u < -2$, $-2 < u < -1$, $-1 < u <0$, and $0 < u < +\infty$, as follows:
\begin{equation}
\begin{split}
&-\infty < u < -2 \quad v \to 0^+: \\
&\xi = \pi \quad \phi = \pi \quad \theta = \pi \quad R = \text{Re}[R_{\text{PV}}(u, 0)] + 0 i,
\end{split}
\end{equation}
\begin{equation}
\begin{split}
&-\infty < u < -2 \quad v \to 0^-: \\
&\xi = \pi \quad \phi = \pi \quad \theta = \pi \quad R = \text{Re}[R_{\text{PV}}(u, 0)] + 0 i,
\end{split}
\end{equation}
\begin{equation}
\begin{split}
&-2 < u < -1 \quad v \to 0^+: \\
&\xi = 0 \quad \phi = \pi \quad \theta = \pi \quad R = \text{Re}[R_{\text{PV}}(u, 0)] - i \pi (2 + u),
\end{split}
\end{equation}
\begin{equation}
\begin{split}
&-2 < u < -1 \quad v \to 0^-: \\
&\xi = 2 \pi \quad \phi = \pi \quad \theta = \pi \quad R = \text{Re}[R_{\text{PV}}(u, 0)] + i \pi (2 + u),
\end{split}
\end{equation}
\begin{equation}
\begin{split}
&-1 < u < 0 \quad v \to 0^+: \\
&\xi = 0 \quad \phi = \pi \quad \theta = 0 \quad R = \text{Re}[R_{\text{PV}}(u, 0)] + i \pi u,
\end{split}
\label{JJ}
\end{equation}
\begin{equation}
\begin{split}
&-1 < u < 0 \quad v \to 0^-: \\
&\xi = 2 \pi \quad \phi = \pi \quad \theta = 2 \pi \quad R = \text{Re}[R_{\text{PV}}(u, 0)] - i \pi u,
\end{split}
\end{equation}
\begin{equation}
\begin{split}
&0 < u < +\infty \quad v \to 0^+: \\
&\xi = 0 \quad \phi = 0 \quad \theta = 0 \quad R = \text{Re}[R_{\text{PV}}(u, 0)] + 0 i,
\end{split}
\label{111}
\end{equation}
\begin{equation}
\begin{split}
&0 < u < +\infty \quad v \to 0^-: \\
&\xi = 2 \pi \quad \phi = 2 \pi \quad \theta = 2 \pi \quad R = \text{Re}[R_{\text{PV}}(u, 0)] + 0 i,
\end{split}
\label{222}
\end{equation}
where $\text{Re}[R_{\text{PV}}(u, v)]$ is given in (\ref{GI}). We notice that this choice of the individual branch cuts all running from the various branch points to $+\infty$ leads to an $R_{\text{PV}}$ that is continuous along the $u$-axis for $-\infty < u < -2$ and then again for $0 < u < +\infty$, but discontinuous in the interval $-2 < u <0$, which identifies the branch cut for $R_{\text{PV}}(w)$ along the $u$-axis.

We have two options regarding the second sheet of the Riemann surface of $R_{\text{PV}}(w)$. The first way to move to the second sheet is through passing the piece of the branch cut from $-1 \to 0$. In this case the angles are: $-2 \pi < \xi < 0$, $0 < \phi < 2 \pi$, and $-2 \pi < \theta < 0$. The another alternative for the second sheet is defined by crossing the piece of the branch cut from $-2 \to -1$, for which the angles now read: $-2 \pi < \xi < 0$, $0 < \phi < 2 \pi$, and $0 < \theta < 2 \pi$. For the sake of concreteness, we choose the first option. Then we have:
\begin{equation}
\begin{split}
&-\infty < u < -2 \quad v \to 0^+: \\
&\xi = -\pi \quad \phi = \pi \quad \theta = -\pi \quad R = \text{Re}[R_{\text{PV}}(u, 0)] + 2 \pi i u,
\end{split}
\end{equation}
\begin{equation}
\begin{split}
&-\infty < u < -2 \quad v \to 0^-: \\
&\xi = -\pi \quad \phi = \pi \quad \theta = -\pi \quad R = \text{Re}[R_{\text{PV}}(u, 0)] + 2 \pi i u,
\end{split}
\end{equation}
\begin{equation}
\begin{split}
&-2 < u < -1 \quad v \to 0^+: \\
&\xi = -2 \pi \quad \phi = \pi \quad \theta = -\pi \quad R = \text{Re}[R_{\text{PV}}(u, 0)] - i \pi (2 - u),
\end{split}
\end{equation}
\begin{equation}
\begin{split}
&-2 < u < -1 \quad v \to 0^-: \\
&\xi = 0 \quad \phi = \pi \quad \theta = -\pi \quad R = \text{Re}[R_{\text{PV}}(u, 0)] + i \pi (2 + 3 u),
\end{split}
\end{equation}
\begin{equation}
\begin{split}
&-1 < u < 0 \quad v \to 0^+: \\
&\xi = -2 \pi \quad \phi = \pi \quad \theta = -2 \pi \quad R = \text{Re}[R_{\text{PV}}(u, 0)] + 3 \pi i u,
\end{split}
\end{equation}
\begin{equation}
\begin{split}
&-1 < u < 0 \quad v \to 0^-: \\
&\xi = 0 \quad \phi = \pi \quad \theta = 0 \quad R = \text{Re}[R_{\text{PV}}(u, 0)] + i \pi u,
\end{split}
\label{JJ1}
\end{equation}
\begin{equation}
\begin{split}
&0 < u < +\infty \quad v \to 0^+: \\
&\xi = -2 \pi \quad \phi = 0 \quad \theta = -2 \pi \quad R = \text{Re}[R_{\text{PV}}(u, 0)] + 2 \pi i u,
\end{split}
\end{equation}
\begin{equation}
\begin{split}
&0 < u < +\infty \quad v \to 0^-: \\
&\xi = 0 \quad \phi = 2 \pi \quad \theta = 0 \quad R = \text{Re}[R_{\text{PV}}(u, 0)] + 2 \pi i u.
\end{split}
\end{equation}
We notice the continuous join of the first sheet upper-lip value of $R_{\text{PV}}(w)$ along the piece of the branch cut from $-1 \to 0$, i.e., (\ref{JJ}), with the second sheet lower-lip value of $R_{\text{PV}}(w)$, that is, (\ref{JJ1}), as we move continuously from positive to negative values of $v$.

\begin{figure}
\centering
\includegraphics[scale = 0.34]{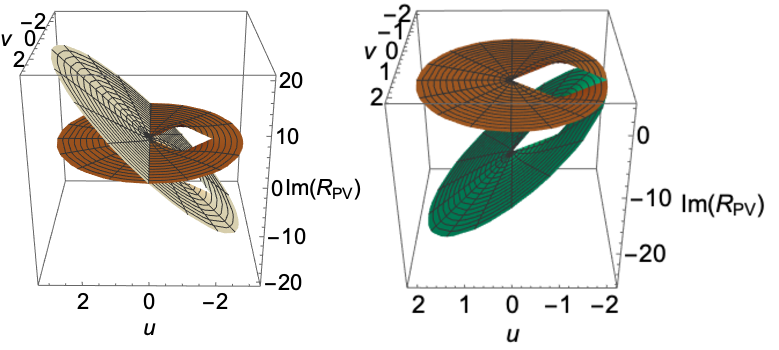}
\caption{The Riemann surface of the complex function $R_{\text{PV}}(w)$. In the left panel, the first two sheets are connected along the piece of the branch cut from $-1 \to 0$ along the $u$-axis; in this case, the angles on the second sheet are: $-2 \pi < \xi < 0$, $0 < \phi < 2 \pi$, and $-2 \pi < \theta < 0$. In the right panel, the sheets are connected along the piece of the branch cut from $-2 \to -1$; now, the angles read: $-2 \pi < \xi < 0$, $0 < \phi < 2 \pi$, and $0 < \theta < 2 \pi$.}
\label{rsmmm}
\end{figure}

In Fig.~\ref{rsmmm}, the two alternatives for the second sheet of $R_{\text{PV}}(w)$ are illustrated. In the left panel, which we discussed above in detail, we move to the second sheet (light brown) by crossing the piece of the branch cut from $-1 \to 0$. The right panel demonstrates the second sheet defined through the angles: $-2 \pi < \xi < 0$, $0 < \phi < 2 \pi$, and $0 < \theta < 2 \pi$, where switching between the sheets takes place by crossing the piece of the branch cut from $-2 \to -1$.

\subsubsection{Continuous quantum phase transition}

For the usual choices of the regulatory cut-off and the coupling strength $\Lambda = 859$ MeV and $G \Lambda^2 = 2.84$ \cite{SPK1}, the applicable equation for obtaining the dynamically generated mass of the fermion is (\ref{111}) or (\ref{222}). We obtain
\begin{equation}
C = 1.1584 = (2 + u) \ln|2 + u| + u \ln|u| - 2 (1 + u) \ln|1 + u|.
\end{equation}
We find the root of this equation at $u_r = 0.0783$; thus the mass is $m^* = 240.367$ MeV.

The strong-coupling regime is defined through $C < R_{\text{PV}}(0) = 2 \ln2$, where $R_{\text{PV}}(0)$ is the maximum value of $R_{\text{PV}}(w)$ for $0 < u < +\infty$.

\begin{figure}
\centering
\includegraphics[scale = 0.6]{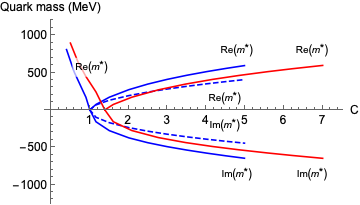}
\caption{The quark mass versus $C$ for both the covariant and Pauli-Villars regularization schemes. The blue curves correspond to the covariant regularization scheme and the red one to the Pauli-Villars. Both regularization schemes show the same response to the variation of $C$, that is, the dynamically generated mass of the quark decreases as $C$ increases, until reaches a critical value, which beyond that value, a continuous quantum phase transition, characterized by the development of a width for the mass, occurs on the second sheet of the Riemann surface. The effect of spiraling down from the second to the third sheet in the case of the covariant regularization scheme, as a dashed curve, is also shown. On this sheet, the angles are considered as $0 < \phi < 2 \pi$ and $-4 \pi < \theta < -2 \pi$.}
\label{figure2}
\end{figure}

As in Sec.~\ref{ss2}, we treat the coupling strength as a parameter and by decreasing its value (and therefore increasing $C$), the dynamically generated fermion mass decreases until the coupling reaches a critical value, $C_c = 2 \ln2$; by going beyond this point and crossing the branch cut in region $-1 \to 0$ onto the second sheet of the Riemann surface of $R_{\text{PV}}(w)$, a continuous quantum phase transition occurs that is characterized by the development of a non-zero width for the mass, see Fig.~\ref{figure2} and Table~\ref{table2}.

\begin{center}
\begin{table}
\begin{tabular}{|c|c|c|c|c|}
\hline
$G \Lambda^2$ & $C$ & $u_r$ & $v_r$ & $m^* = m - i \gamma$\\
\colrule
$2.37$ & $1.39$ & $0.0000$ & $0.0000$ & $0.0000 - 0.0000 i$\\

$2.05$ & $1.60$ & $-0.0269$ & $-0.0311$ & $72.4305 - 158.4150 i$\\

$1.58$ & $2.08$ & $-0.0648$ & $-0.1273$ & $169.6860 - 276.7820 i$\\

$1.19$ & $2.76$ & $-0.0890$ & $-0.2698$ & $268.2920 - 371.0150 i$\\

$0.95$ & $3.46$ & $-0.0981$ & $-0.4094$ & $345.1480 - 437.6220 i$\\

$0.79$ & $4.16$ & $-0.1004$ & $-0.5434$ & $408.4530 - 490.8330 i$\\

$0.68$ & $4.84$ & $-0.0995$ & $-0.6676$ & $460.7770 - 534.5420 i$\\

$0.59$ & $5.58$ & $-0.0969$ & $-0.7999$ & $511.4290 - 577.1130 i$\\

$0.53$ & $6.21$ & $-0.0941$ & $-0.9107$ & $550.5140 - 610.3280 i$\\

$0.47$ & $7.00$ & $-0.0901$ & $-1.0475$ & $595.5260 - 648.9480 i$\\
\hline
\end{tabular}
\caption{Quark mass as a function of $C$, in the weak-coupling regime, $C \geq 2 \ln2$, on the second sheet of the Riemann surface of $R_{\text{PV}}(w)$. The mass scale has been set using $\Lambda = 859$ MeV.}
\label{table2}
\end{table}
\end{center}
Thus, as was the case with the covariant regularization scheme, the gap equation and the resulting generated mass respond in the same manner to the variation of the coupling strength.

\begin{figure}
\centering
\includegraphics[scale = 0.6]{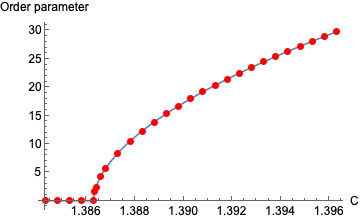}
\caption{In the vicinity of the phase transition point, i.e., $2 \ln2$, the order parameter behaves as a power law with the same critical exponent as obtained in \ref{ss2}: $\text{Order parameter} \propto (C - 2 \ln2)^{0.55}$. The blue curve is fitted to the numerical data as red dots.}
\label{rpvop}
\end{figure}

Further, we note that the order parameter diverges as a power law with the same critical exponent as was obtained in the case of the covariant regularization scheme, that is, |Im($m^*$)| $ \propto (C - 2 \ln2)^\beta$, where once again $\beta \approx 0.55$, see Fig.~\ref{rpvop}.

%------------------------------------------
\subsection{Solutions of the gap equation in the proper-time regularization scheme and the effects of a constant electric field on the dynamically generated mass}
\label{ss4}

To study the effects of external electromagnetic fields on the system, it is convenient to use the proper-time regularization scheme. Then, the gap equation reads
\begin{equation}
m^* =  \frac{1}{2 \pi^2} N_c N_f G m^* \int_{1 / \Lambda^2}^{+\infty} \frac{ds}{s^2} e^{- {m^*}^2 s}.
\label{ptg}
\end{equation}
This becomes
\begin{equation}
2 \pi^2 / (N_c N_f G \Lambda^2) = e^{- z^2} - z^2 \Gamma(0, z^2),
\label{popo}
\end{equation}
where $z = m^* / \Lambda$ and the incomplete Gamma functions are defined as
\begin{equation}
\gamma(\alpha, z) = \int_{0}^{z} dt \, e^{-t} t^{\alpha - 1} \quad \quad \Gamma(\alpha, z) = \int_{z}^{+\infty} dt \, e^{-t} t^{\alpha - 1}.
\end{equation}
As was the case with the covariant regularization scheme in Sec.~\ref{ss2}, the right-hand side of (\ref{popo}) has a global maximum of $1$, so the real solution of the gap equation lies in the strong-coupling regime: $2 \pi^2 / (N_c N_f) < G \Lambda^2$. This condition is satisfied by the choice of parameters given in \cite{SPK1} as $\Lambda = 1086$ MeV and $G \Lambda^2 = 3.78$ which result in $m^* = 199.987$ MeV.

\begin{figure}
\centering
\includegraphics[scale = 0.34]{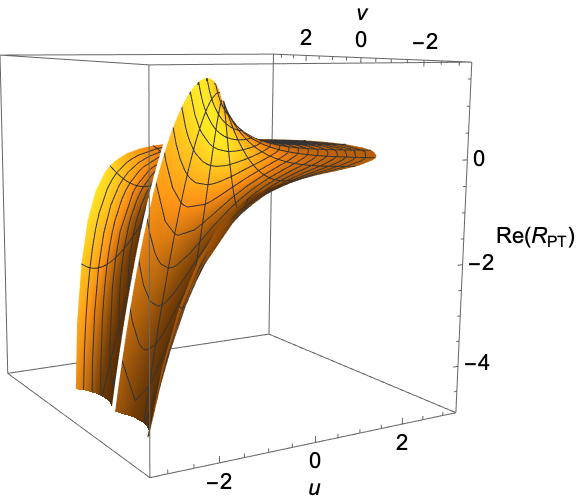}
\caption{The branch cut of the complex function $R_{\text{PT}}(w)$ is introduced on the negative $u$-axis (white line).}
\label{rptp}
\end{figure}

In order to treat the coupling strength as a variable, we consider the complex version of (\ref{popo}), which reads
\begin{equation}
2 \pi^2 / (N_c N_f G \Lambda^2) = R_{\text{PT}}(w) = e^{-w} - w \Gamma(0, w),
\end{equation}
in terms of the variable $z^2 = w = u + i v$. $R_{\text{PT}}(w)$ has a branch point at the origin and we introduce the branch cut along the negative $u$-axis, see Fig.~\ref{rptp}.

\begin{figure}
\centering
\includegraphics[scale = 0.4]{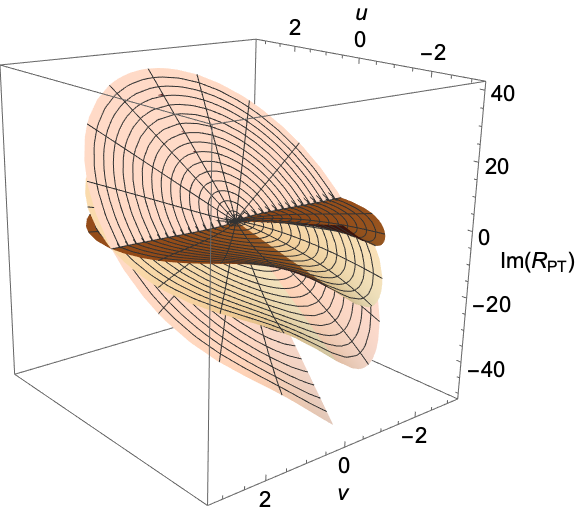}
\caption{The Riemann surface of the complex function $R_{\text{PT}}(w)$. By encircling the origin, we move to other sheets of the Riemann surface.}
\label{ript}
\end{figure}

To construct the Riemann surface of $R_{\text{PT}}(w)$, first we note that
\begin{equation}
\gamma(\alpha, w e^{2 k \pi i}) = e^{2 k \alpha \pi i} \gamma(\alpha, w) \quad \quad \gamma(\alpha, w) + \Gamma(\alpha, w) = \Gamma(\alpha).
\end{equation}
From these relations, we obtain
\begin{equation}
\Gamma(\alpha, w e^{2 k \pi i}) = e^{2 k \alpha \pi i} \Gamma(\alpha, w) + (1 - e^{2 k \alpha \pi i}) \Gamma(\alpha),
\label{gg}
\end{equation}
for some integer $k$. The limiting value of (\ref{gg}) as $\alpha \to 0$ is
\begin{equation}
\Gamma(0, w e^{2 k \pi i}) = \Gamma(0, w) - 2 k \pi i,
\end{equation}
where we have used $\lim_{\alpha \to 0} \alpha \Gamma(\alpha) = 1$.
Thus we have
\begin{equation}
R_{\text{PT}}(w) = e^{-w} - w \Gamma(0, w) + 2 k \pi i w.
\end{equation}
For different values of $k$ we obtain the different sheets of $R_{\text{PT}}(w)$, with $k = 0$ corresponding to the first sheet. In Fig.~\ref{ript}, we show the first three sheets of the Riemann surface of $R_{\text{PT}}(w)$, which are joined along the negative $u$-axis. By encircling the origin one (two) time(s) we move to the second (third) sheet of $R_{\text{PT}}(w)$.

By decreasing the coupling strength the dynamically generated fermion mass that lies on the first sheet ($k = 0$) of the Riemann surface of $R_{\text{PT}}(w)$, decreases until reaches a phase transition point at $C_c = 2 \pi^2 / (N_c N_f G_c \Lambda^2) = 1$; then, by encircling the origin and moving to the second sheet ($k = 1$) of $R_{\text{PT}}(w)$, the mass develops an imaginary part in the weak-coupling region, see Table~\ref{table111}.
\begin{center}
\begin{table}
\begin{tabular}{|c|c|c|c|c|}
\hline
$G \Lambda^2$ & $C$ & $u_r$ & $v_r$ & $m^* = m - i \gamma$\\
\colrule
$3.78$ & $0.87$ & $0.0339$ & $0.0000$ & $199.9870 - 0.0000 i$\\

$3.60$ & $0.91$ & $0.0198$ & $0.0000$ & $152.7270 - 0.0000 i$\\

$3.45$ & $0.95$ & $0.0090$ & $0.0000$ & $103.2680 - 0.0000 i$\\

$3.40$ & $0.97$ & $0.0058$ & $0.0000$ & $82.7904 - 0.0000 i$\\

$3.33$ & $0.99$ & $0.0018$ & $0.0000$ & $45.8811 - 0.0000 i$\\

$\pi^2 / 3$ & $1.00$ & $0.0000$ & $0.0000$ & $0.0000 - 0.0000 i$\\

$2.84$ & $1.16$ & $-0.0201$ & $-0.0199$ & $69.4623 - 169.0670 i$\\

$2.19$ & $1.50$ & $-0.0546$ & $-0.0849$ & $165.3940 - 302.8690 i$\\

$1.64$ & $2.01$ & $-0.0848$ & $-0.1914$ & $270.9980 - 416.4820 i$\\

$1.32$ & $2.49$ & $-0.1005$ & $-0.2948$ & $352.6970 - 492.8250 i$\\

$1.10$ & $2.99$ & $-0.1084$ & $-0.3984$ & $423.7320 - 554.4520 i$\\

$0.94$ & $3.50$ & $-0.1112$ & $-0.5013$ & $487.0580 - 606.9840 i$\\

$0.82$ & $4.01$ & $-0.1106$ & $-0.6020$ & $543.8210 - 652.8040 i$\\

$0.73$ & $4.51$ & $-0.1077$ & $-0.6968$ & $593.5180 - 692.2890 i$\\

$0.66$ & $4.98$ & $-0.1034$ & $-0.7863$ & $637.6940 - 727.0950 i$\\
\hline
\end{tabular}
\caption{Quark mass decreases as the coupling strength decreases, until reaches a phase transition point, where the mass gains a width by moving to the weak-coupling regime of the theory onto the second sheet of the Riemann surface of $R_{\text{PT}}(w)$. The mass scale is $\Lambda = 1086$ MeV.}
\label{table111}
\end{table}
\end{center}

The power-law behavior of the order parameter in the vicinity of the phase transition point is defined as before: |Im($m^*$)| $\propto (C - 1)^\beta$, and once again we find that $\beta \approx 0.55$. This behavior is the same in the other two regularization schemes in Secs.~\ref{ss2} and \ref{ss3}, which establishes the fact that the power-law divergence of the order parameter close to the phase transition point with the critical exponent $\beta \approx 0.55$ is independent of the choice of the regularization scheme and the particular form of the resulting gap equation.

In the presence of a constant electric field, the gap equation (\ref{ptg}) takes the form
\begin{equation}
\begin{split}
m^* = &\frac{1}{2 \pi^2} N_c G m^* \bigg[ N_f \int_{1 / \Lambda^2}^{+\infty} \frac{ds}{s^2} e^{- {m^*}^2 s}\\
&+ \sum_{f} \int_{0}^{+\infty} \frac{ds}{s^2} e^{- {m^*}^2 s} [q_f E s \cot(q_f E s) - 1] \bigg],
\end{split}
\label{horr}
\end{equation}
where $m^*$ is the field-dependent dynamical mass, $m^*=m^*(E)$. In (\ref{horr}) we have split up the gap equation so as to isolate the divergence in the first term.

To calculate the second term on the right-hand side of (\ref{horr}), we first note that
\begin{equation}
\begin{split}
&\int_{0}^{+\infty} \frac{ds}{s^2} e^{- {m^*}^2 s} [q_f E s \cot(q_f E s) - 1] =\\
&(2 q_f E) (-2) \int_{0}^{+\infty} ds \frac{\tanh^{-1} (2 q_f E s / {m^*}^2)}{e^{2 \pi s} - 1}.
\end{split}
\label{rruuoo}
\end{equation}
By exploiting Binet's second expression for $\ln \Gamma(z)$ \cite{gaw},
\begin{equation}
\begin{split}
\ln \Gamma(z) = &(z - 1/2) \ln z - z + (1 / 2) \ln 2\pi\\
&+ 2 \int_{0}^{+\infty} ds \frac{\tan^{-1} (s / z)}{e^{2 \pi s} - 1},
\end{split}
\end{equation}
we can rewrite (\ref{rruuoo}) as
\begin{equation}
\begin{split}
\int_{0}^{+\infty} \frac{ds}{s^2} e^{- {m^*}^2 s} &[q_f E s \cot(q_f E s) - 1] =\\
&q_f E \, \text{Re} \big[ J[i {m^*}^2 / (2 q_f E)] \big],
\end{split}
\label{horrop}
\end{equation}
where
\begin{equation}
J(z) = 2 i [(z - 1/2) \ln z - z + (1 / 2) \ln 2 \pi - \ln \Gamma(z)].
\end{equation}
By assuming no charge difference for the fermions and denoting $q_f E$ as $a$, $q_f E J[i {m^*}^2 / (2 q_f E)]$ can be written as
\begin{equation}
\begin{split}
a J[i {m^*}^2 / (2 a)] = &{m^*}^2 + i a \ln 2 \pi - ({m^*}^2 + i a) \ln [i {m^*}^2 / (2 a)]\\
&- 2 i a \ln \Gamma[i {m^*}^2 / (2 a)].
\end{split}
\label{hhhrrr}
\end{equation}
Then, to obtain the real part of (\ref{hhhrrr}) as required in (\ref{horrop}), we note that the quantity ${m^*}^2 / (2 a)$ is positive, thus $\ln [i {m^*}^2 / (2 a)]$ becomes
\begin{equation}
\ln (i b) = \ln b + i \pi / 2,
\end{equation}
where $b \equiv {m^*}^2 / (2 a)$.

In order to treat the function $\ln \Gamma[i {m^*}^2 / (2 a)]$, we use the approximation \cite{pcm},
\begin{equation}
\begin{split}
\ln \Gamma(i b) \approx &- i b^{-1} / 12 - i b + i b \ln (i b)\\
&- (1 / 2) \ln (i b) + (1 / 2) \ln 2 \pi,
\end{split}
\end{equation}
which is accurate for $b \geq 1$. Now, the right-hand side of (\ref{horrop}) becomes
\begin{equation}
a \text{Re} [ J (i b) ] = - a b^{-1} / 6.
\end{equation}
Finally, the gap equation in the presence of a constant electric field, Eq.~(\ref{horr}), reduces to
\begin{equation}
2 \pi^2 / (N_c N_f G \Lambda^2) = e^{- z^2} - z^2 \Gamma(0, z^2) - \frac{1}{3} \frac{(Q {\cal E})^2}{z^2},
\end{equation}
where $z = m^* / \Lambda$, $Q = q / \Lambda$, and ${\cal E} = E / \Lambda$.

For small values of $b$, we approximate $\ln \Gamma(i b)$ as
\begin{equation}
\ln \Gamma(i b) \approx - i b \gamma_{EM} - b^2 \pi^2 / 12 - \ln (i b),
\end{equation}
where $\gamma_{EM}$ is the Euler-Mascheroni constant. Then the gap equation (\ref{horr}) becomes
\begin{equation}
\begin{split}
2 \pi^2 / (N_c N_f G \Lambda^2) = e^{- z^2} &- z^2 \Big[ \Gamma(0, z^2) + \ln [ z^2 / (2 Q {\cal E}) ]\\
&+ \gamma_{EM} - 1 \Big] - Q {\cal E} \pi / 2.
\label{smallb}
\end{split}
\end{equation}
The limiting value of the right-hand side of (\ref{smallb}) as $z \to 0$ is $1 - Q {\cal E} \pi / 2$. Thus, in the presence of a constant electric field the critical value of the coupling strength reads $G_c \Lambda^2 = 2 \pi^2 / [N_c N_f (1 - Q {\cal E} \pi / 2)]$. We note that when the electric field vanishes, we recover the relation $2 \pi^2 / (N_c N_f) < G \Lambda^2$.

\begin{figure}
\centering
\includegraphics[scale = 0.36]{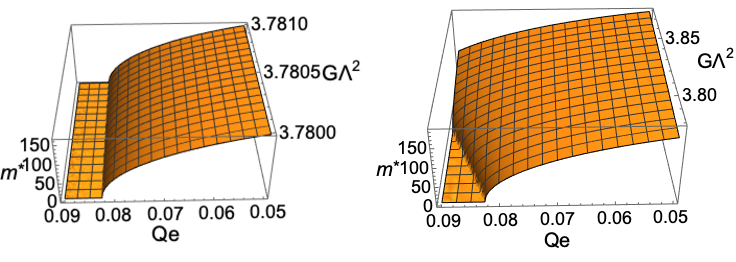}
\caption{The left panel shows how the dynamically generated mass decreases as the electric field increases. This effect of the electric field is compensated by increasing the coupling strength, as it is illustrated in the right panel. ($Qe$ denotes $Q{\cal E}$.)}
\label{ptfigpt}
\end{figure}

By fixing $\Lambda = 1086$ MeV and $G \Lambda^2 = 3.78$, by starting from $Q {\cal E} = 0.05$ and gradually increasing it, we observe that the dynamically generated mass decreases until $Q {\cal E}$ reaches a critical point: ${Q {\cal E}}_c = (2 / \pi) [1 - 2 \pi^2 / (N_c N_f G \Lambda^2)] \approx 0.08255$, where $m^*$ vanishes. This is illustrated in the left panel of Fig.~\ref{ptfigpt}; the right panel shows how increasing the coupling strength compensates the effect of increasing the electric field.

\begin{figure}
\centering
\includegraphics[scale = 0.4]{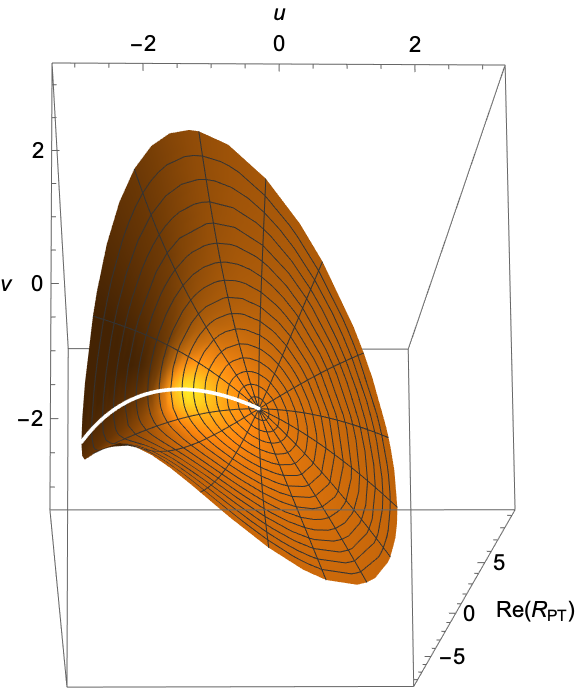}
\caption{The picture illustrates the gap equation in the presence of a constant electric field. The branch cut is introduced on the negative $u$-axis (white line).}
\label{elecpt}
\end{figure}

We now fix $G \Lambda^2 = 3.78$ and treat the electric field as a variable which can be increased arbitrarily. To this end, our complex gap equation reads
\begin{equation}
\begin{split}
2 \pi^2 / (N_c N_f G \Lambda^2) = e^{- w} &- w \Big[ \Gamma(0, w) + \ln [ w / (2 Q {\cal E}) ]\\
&+ \gamma_{EM} - 1 \Big] - Q {\cal E} \pi / 2,
\end{split}
\label{smallbbb}
\end{equation}
where $z^2 = w = u + i v$. For a fixed $Q {\cal E} = 0.07$, we show the right-hand side of (\ref{smallbbb}) in Fig.~\ref{elecpt}. This is to be compared with Fig.~\ref{rptp}, where the gap equation is obtained in the absence of a medium.

By increasing the electric field, the dynamical mass decreases until it vanishes at a phase transition point and develops a negative imaginary part by moving to the second sheet of the Riemann surface associated with the complex gap equation; this is demonstrated in Table~\ref{table00pt}. We note that by increasing the electric field the imaginary part of the mass also increases.
\begin{center}
\begin{table}
\begin{tabular}{|c|c|}
\hline
$Q {\cal E}$ & $m^* = m - i \gamma$\\
\colrule
$0.050$ & $161.4360 - 0.0000 i$\\

$0.055$ & $151.7200 - 0.0000 i$\\

$0.060$ & $140.0850 - 0.0000 i$\\

$0.065$ & $126.0200 - 0.0000 i$\\

$0.070$ & $108.5940 - 0.0000 i$\\

$0.075$ & $85.7774 - 0.0000 i$\\

$0.080$ & $50.7274 - 0.0000 i$\\

$0.082$ & $23.6737 - 0.0000 i$\\

$(2 / \pi) [1 - 2 \pi^2 / (N_c N_f G \Lambda^2)]$ & $0.0000 - 0.0000 i$\\

$0.085$ & $15.9243 - 21.0358 i$\\

$0.090$ & $27.9443 - 36.5902 i$\\

$0.095$ & $36.3496 - 47.2000 i$\\

$0.100$ & $43.2883 - 55.7652 i$\\

$0.105$ & $49.3741 - 63.1254 i$\\

$0.110$ & $54.8848 - 69.6654 i$\\

$0.115$ & $59.9736 - 75.5996 i$\\
\hline
\end{tabular}
\caption{The dynamically generated quark mass as a function of the electric field. The mass decreases as the electric field increases, until it vanishes at a phase transition point, i.e., ${Q {\cal E}}_c = (2 / \pi) [1 - 2 \pi^2 / (N_c N_f G \Lambda^2)] \approx 0.08255$, and then develops an imaginary part by moving to the second sheet of the Riemann surface. The coupling strength is fixed as $G \Lambda^2 = 3.78$.}
\label{table00pt}
\end{table}
\end{center}

\section{Meson masses}
\label{s4}

\subsection{Pseudoscalar sector}
\label{s41}

The mass of the isovector pseudoscalar mode that corresponds to the $\pi$ meson invoked by the interaction term $(\bar\psi i \gamma_5 \tau \psi)^2$ in the NJL Hamiltonian (\ref{eq:2}) is determined by computing the effective scattering amplitude or effective exchange interaction, which can be expressed as a geometric sum of proper polarization graphs $\Pi_{\text{ps}}(k^2)$.  This effective interaction is proportional to $1/[1 - 2 G \Pi_{\text{ps}}(k^2)]$, so that the poles of this expression correspond to the pseudoscalar mode that is excited. 

Quite generally, the pseudoscalar proper polarization is given as
\begin{equation}
\begin{split}
\frac{1}{i} \Pi_{\text{ps}}(k^2) = -\int \frac{d^4 p}{(2 \pi)^4} &\mathrm{Tr} i \gamma_5 T_i i S(p + k / 2)\\
&\times i \gamma_5 T_j i S(p - k / 2),
\end{split}
\label{ver}
\end{equation}
where $T$ selects the isospin channel for creating a $\pi$ meson and $S(p)$ is given in (\ref{fey1}). By performing the trace on color, spinor, and flavor indices, the above reduces to
\begin{equation}
\begin{split}
\frac{1}{i} \Pi_{\text{ps}}(k^2) = -4 N_c N_f \int \frac{d^4 p}{(2 \pi)^4} &\frac{(m - i \gamma)^2 - p^2 + \frac{1}{4} k^2}{[(p + \frac{1}{2} k)^2 - (m - i \gamma)^2]}\\
&\times \frac{1}{[(p - \frac{1}{2} k)^2 - (m - i \gamma)^2]}.
\end{split}
\end{equation}
Rewriting the denominator in terms of partial fractions and making suitable shifts of variables, the above can be written as
\begin{equation}
\begin{split}
\frac{1}{i} \Pi_{\text{ps}}(k^2) = &4 N_c N_f \int \frac{d^4 p}{(2 \pi)^4} \frac{1}{p^2 - (m - i \gamma)^2}\\
&-2 N_c N_f k^2 I(k^2),
\end{split}
\label{iii}
\end{equation}
where
\begin{equation}
\begin{split}
I(k^2) = \int \frac{d^4 p}{(2 \pi)^4} &\frac{1}{[(p + \frac{1}{2} k)^2 - (m - i \gamma)^2]}\\
&\times \frac{1}{[(p - \frac{1}{2} k)^2 - (m - i \gamma)^2]}.
\end{split}
\end{equation}
By exploiting the gap equation, one can eliminate the integral in (\ref{iii}), and obtain
\begin{equation}
1 - 2 G \Pi_{\text{ps}}(k^2) = 4 i N_c N_f G k^2 I(k^2),
\end{equation}
which still has a real root, corresponding to a real pseudoscalar mass, when $k^2 = 0$. Thus, the Goldstone mode is impervious to the possible complex nature of the constituent masses obtained as a result of the continuous quantum phase transition. This result is independent of the regularization procedure employed.

\subsection{Scalar sector}
\label{s42}

To calculate the mass of the isoscalar scalar mode, corresponding to the $\sigma$ meson (usually associated with the term $(\bar\psi \psi)^2$ of (\ref{eq:2})), we compute the scalar proper polarization 
\begin{equation}
\frac{1}{i} \Pi_{\text{s}}(k^2) = -\int \frac{d^4 p}{(2 \pi)^4} \mathrm{Tr} i S(p + k / 2) i S(p - k / 2),
\end{equation}
where we have replaced the vertex factor of $i \gamma_5 T$ in (\ref{ver}) by $1$, in both spinor and flavor space. By performing the trace we obtain:
\begin{equation}
\begin{split}
\frac{1}{i} \Pi_{\text{s}}(k^2) = &4 N_c N_f \int \frac{d^4 p}{(2 \pi)^4} \frac{1}{p^2 - (m - i \gamma)^2}\\
&-2 N_c N_f [k^2 - 4 (m - i \gamma)^2] I(k^2).
\end{split}
\end{equation}
Making use of the gap equation and rearranging terms, it follows that
\begin{equation}
1 - 2 G \Pi_{\text{s}}(k^2) = 4 i N_c N_f G [k^2 - 4 (m - i \gamma)^2] I(k^2).
\end{equation}
From this, we obtain the mass of the scalar meson to be
\begin{equation}
m_\sigma = \pm 2 (m - i \gamma).
\end{equation}
Thus the dynamical generation of a width for the fermion mass causes the scalar meson mass also to gain a width.

\section{Concluding remarks}
\label{s6}

We have investigated the dynamical generation of mass as a function of the coupling strength $G$ of the NJL model; we have kept the notation of its use as a strong-coupling model for quantum chromodynamics, in order to check the numerical values that we obtain. However, we consider the results as a playground for observing dynamical symmetry breaking for systems with two fermion species having appropriate couplings. 

In our investigation of the behavior of the dynamically generated fermion mass, we observe a continuous quantum phase transition characterized by the generation of a width on the higher sheets of the Riemann surface associated with the gap equation, when the interaction strength falls below a critical value, i.e., in the weak-coupling regime. In the vicinity of the phase transition point, we find a power-law behavior with a critical exponent, $\beta \approx 0.55$; this is found to be independent of the choice of the regularization scheme.

In the weak-coupling regime, the dynamically generated fermion mass takes on a complex structure as $m^* = m - i \gamma$ on the second sheet of the Riemann surface. In other words, the imaginary part of the mass is always negative. This implies that the fermion dressing that gives it mass is unstable, that is, the coupling is too weak to dress the fermion permanently and the states of the system can only decay in time. (This situation resembles the eigenvalues and time asymmetry of an open quantum system.) One possible implication of this could be that the chirally broken vacuum decays back into a chiral conserving vacuum by  emitting a Goldstone boson in the process.

Energetically, in the regime in which the coupling strength falls below the critical value, the normal vacuum is favored over the condensed one, contrary to the strong-coupling regime. This can be seen in the formula obtained in Appendix C of \cite{SPK1}, where the change in energy density between the condensed and normal phases is found to be:
\begin{equation}
\begin{split}
\langle \delta T^{00} \rangle &= \langle T^{00} \rangle_{\text{condensed}} - \langle T^{00} \rangle_{\text{normal}}\\
&\propto - \Big(1 - \frac {G_c \Lambda^2} {G \Lambda^2}\Big).
\end{split}
\end{equation}
The right-hand side of the above is positive in the weak-coupling regime, indicating that the normal phase is the energetically favored state. This implies that the anomalous states found here can only be accessed by a driven process.

A similar continuous quantum phase transition can also be obtained by fixing the coupling strength to its strong value, and introducing an external parameter such as a constant electric field: On increasing this field beyond its critical value, the dynamically generated mass develops a negative imaginary part when moving to the second sheet of the Riemann surface.

Accompanying the appearance of a width for the dynamically generated fermion mass as a continuous function of the system parameter, we find that the behavior of the isovector pseudoscalar mode is unchanged, that is, it remains a Goldstone boson and has zero mass. On the other hand, the behavior of the isoscalar scalar particle follows that of the order parameter itself and gains a width.

These results suggest that in similar cases in the NJL model where a phase transition occurs, a similar behavior can be expected. Thus, for example, in the case in which the coupling strength is held fixed,
but temperature is varied, the response of the order parameter, and thus the mass of the scalar particle, will be to gain a width.

We conclude by commenting that it is only very recently that experiments that make use of the Riemann surface structure of the complex functions have been able to be performed. Encircling a branch point and switching between sheets of the Riemann surface have been performed experimentally in different areas of physics: In \cite{mmmppp}, the authors have demonstrated the transfer of energy between two states of the system, which arises from the presence of a branch point in the spectrum. In \cite{lloo}, the branch point is fully encircled dynamically, and this has made it possible a robust asymmetric switch between the two sheets of the Riemann surface. We are thus optimistic that it may in the future be possible to study the properties of systems of interacting fermions through quantum or other simulators.

\end{document}